\newcommand\vx{\vec{x}}

\newcommand\deltam{\delta_{\rm m}}

\newcommand\deltagal{\delta_{\rm g}}

\newcommand\omegam{\Omega_{\rm m}}

\newcommand{\specialcell}[2][l]{%
  \begin{tabular}[#1]{@{}l@{}}#2\end{tabular}}

\newcommand\DV{D_{\rm V}}

\newcommand\Mpch{{\rm\; Mpc}/h}

\newcommand{\vbc}{v_{\rm bc}}

\documentclass[useAMS,usenatbib]{mn2e}

\usepackage{lineno}

\usepackage{placeins}

\usepackage{graphicx} 
\usepackage{multirow}
\usepackage{array}

\usepackage{amsmath,amssymb}

\usepackage[T1]{fontenc}
\usepackage[latin9]{inputenc}
\usepackage{float}
\usepackage{graphicx}
\usepackage{esint}

\usepackage{hyperref}

\makeatletter
\DeclareFontEncoding{LGR}{}{}
\DeclareTextSymbol{\~}{LGR}{126}

\begin{document}

\title[Relative velocity in the 3PCF]{Constraining the Baryon-Dark Matter Relative Velocity with the Large-Scale 3-Point Correlation Function of the SDSS BOSS DR12 CMASS Galaxies}

\author{\makeatauthor}
\author[Slepian et al.]{Zachary Slepian$^{1}$\thanks{zslepian@cfa.harvard.edu},
Daniel J. Eisenstein$^{1}$\thanks{deisenstein@cfa.harvard.edu},
Jonathan A. Blazek$^{2}$,
Joel R. Brownstein$^{3}$,\and
Chia-Hsun Chuang$^{4,5}$, 
H\'ector Gil-Mar\'in$^{6, 7}$,
Shirley Ho$^{8,9,10}$,
Francisco-Shu Kitaura$^5$,\and
Joseph E. McEwen$^{2}$,
Will J. Percival$^{11}$,
Ashley J. Ross$^{2}$,
Graziano Rossi$^{12}$,\and
Hee-Jong Seo$^{13}$,
An\v{z}e Slosar$^{14}$
\& Mariana Vargas-Maga\~na$^{15}$\\
$^{1}$ Harvard-Smithsonian Center for Astrophysics, 60 Garden Street, Cambridge, MA 02138, USA\\
$^{2}$ Center for Cosmology and Astroparticle Physics, Department of Physics, The Ohio State University, OH 43210, USA\\
$^{3}$Department of Physics and Astronomy, University of Utah, Salt Lake City, UT 84112, USA\\
$^4$ Instituto de F\'{\i}sica Te\'orica, (UAM/CSIC), Universidad Aut\'onoma de Madrid, Cantoblanco, E-28049 Madrid, Spain \\
$^5$ Leibniz-Institut f\"{u}r Astrophysik Potsdam (AIP), An der Sternwarte 16, D-14482 Potsdam, Germany\\
$^{6}$ Sorbonne Universit\'es, Institut Lagrange de Paris (ILP), 98 bis Boulevard Arago, 75014 Paris, France\\
$^{7}$ Laboratoire de Physique Nucl\'eaire et de Hautes Energies, Universit\'e Pierre et Marie Curie, 4 Place Jussieu, 75005 Paris, France\\
$^8$ Lawrence Berkeley National Lab, 1 Cyclotron Rd, Berkeley CA 94720, USA\\
$^9$ McWilliams Center for Cosmology, Department of Physics, Carnegie Mellon University, 5000 Forbes Ave., Pittsburgh, PA 15213, USA\\
$^{10}$ Department of Physics, University of California, Berkeley, CA 94720, USA\\
$^{11}$ Institute of Cosmology \& Gravitation, University of Portsmouth, Dennis Sciama Building, Portsmouth PO1 3FX, UK\\
$^{12}$ Department of Physics and Astronomy, Sejong University, Seoul, 143-747, Korea\\
$^{13}$ Department of Physics and Astronomy, Ohio University, Clippinger Labs, Athens, OH 45701, USA\\
$^{14}$ Brookhaven National Laboratory, Upton, NY 11973, USA\\
$^{15}$ Instituto de F\'isica, Universidad Nacional Aut\'onoma de M\'exico, Apdo. Postal 20-364, M\'exico}

\maketitle

\begin{abstract}
We search for a galaxy clustering bias due to a modulation of galaxy number with the baryon-dark matter
relative velocity resulting from recombination-era physics.  We find no detected signal and place the constraint $b_v < 0.01$ on the relative velocity bias for the CMASS galaxies. This bias is an important potential systematic of Baryon Acoustic Oscillation (BAO) method measurements of the cosmic distance scale using the 2-point clustering. Our limit on
the relative velocity bias indicates a systematic shift of no more than $0.3\%$ rms in the distance scale inferred from the BAO feature in the BOSS 2-point clustering, well below the $1\%$ statistical error of this measurement. This constraint is the most stringent currently available and has important implications for the ability of upcoming large-scale structure surveys such as DESI to self-protect against the relative velocity as a possible systematic.

\end{abstract}

\section{Introduction}
\label{sec:intro}
Prior to decoupling at redshift $z\sim 1020$, baryons and dark matter behave differently because they experience different forces. The Universe is ionized, and the electrons are tightly coupled to the photons through Thomson scattering, while the protons follow the electrons under the Coulomb force.  On scales within the sound horizon, the baryons are supported against gravitational infall by the photon pressure, which is large because the photons are an important component of the energy density and are relativistic.  

Consider the evolution of a point-like density perturbation in an otherwise homogeneous Universe. It will create a photon overpressure that launches a pulse of baryons and photons outwards (i.e. produce a Baryon Acoustic Oscillation (BAO)), and this pulse's front will be at the sound horizon (Sakharov 1966; Peebles \& Yu 1970; Sunyaev \& Zel'dovich 1970; Bond \& Efstathiou 1984, 1987; Holtzmann 1989; Hu \& Sugiyama 1996; Eisenstein \& Hu 1998; Eisenstein, Seo \& White 2007; Slepian \& Eisenstein 2016a).

Baryons and photons farther away from the overdensity than the sound horizon will not yet know about the overpressure, and so they must infall under gravity. Meanwhile, the dark matter is insensitive to the photon pressure and infalls under gravity on all scales. As a result, when the photons release the baryons at decoupling, the dark matter within the sound horizon has a head start on infalling towards the initial density perturbation: there is a baryon-dark matter relative velocity on scales within the sound horizon.

The magnitude of this relative velocity depends on the magnitude of the initial density perturbation. Therefore different regions of the Universe have different relative velocities, and these velocities are coherent on sound-horizon ($100\Mpch$) scales.  The relative velocity effect was first calculated by Tseliakhovich \& Hirata (2010), and shortly after (Dalal, Pen \& Seljak 2010; Yoo, Dalal \& Seljak 2011) it was shown that this relative velocity can shift the BAO signal in the 2-point clustering if the late-time Luminous Red Galaxies (LRGs) used for these measurements have strong memories of their earliest progenitors.  In particular, the relative velocity's root mean square value at $z\sim 50$, when the first galaxies are expected to form, is of order $10\%$ of the smallest $10^6\;M_{\odot}$ dark matter haloes' circular velocities or velocity dispersions.  Thus small dark matter haloes living in a region of high relative velocity will find it difficult to capture baryons: the baryons' kinetic energy in the dark matter's rest frame is too large. The relative velocity can therefore induce an additional modulation of the clustering of these primordial galaxies on scales out to the BAO scale of $100\Mpch$.  This modulation adds or subtracts from the primordial 2-Point Correlation Function (2PCF) within the BAO scale but not outside, and so can shift the BAO bump in or out in physical scale (this configuration space picture was devloped in Slepian \& Eisenstein 2015a, hereafter SE15a). If the late-time LRGs used for the BAO method at present have a strong memory of their early, small progenitors, the relative velocity can therefore bias the measured cosmic distance scale. Note that the relative velocity is fundamentally an effect set by the relativistic sound speed prior to decoupling; thus its large-scale coherence is unique and cannot be substantially modified by later-time feedback processes as or non-linear structure formation as they operate on far smaller scales. The most recent work on this bias has shown that even a small coupling of the relative velocity to late-time galaxy formation can induce a substantial shift in the distance scale (Blazek, McEwen \& Hirata 2016). 

Yoo, Dalal \& Seljak (2011) proposed that the bispectrum (Fourier space analog of the 3-Point Correlation Function (3PCF)) could be used to measure the relative velocity bias and then correct any effect in the 2PCF, but up to now this technique has not been used.  Yoo \& Seljak (2013) used the power spectrum of galaxies to constrain the relative velocity bias in their bias model to be less than $0.033$; due to different normalization conventions this translates to a $b_v$ constraint in our bias model of $0.1$, as we further detail in \S\ref{sec:concs}. Beutler et al. (2016) compared different redshift slices within the Baryon Oscillation Spectroscopic Survey (BOSS) and WiggleZ to look for the relative velocity effect and found no evidence for it.  Higher precision constraints are required: Blazek, McEwen \& Hirata (2016) show that even a relative velocity bias of order $4\%$ of the linear bias can cause a $1\%$ shift in the distance scale, comparable with the statistical errors on the lastest BOSS measurement (Cuesta et al. 2016; Gil-Mar\'in et al. 2016; Alam et al. 2016).  Further, Dark Energy Spectroscopic Instrument (DESI; Levi et al. 2013), with first light in 2019, will improve on the BOSS error bars by roughly a factor of five, so even a relative velocity bias that is $1\%$ of the linear bias could systematically shift the distance scale comparably to DESI's statistical error bars. Given that any robustly detected deviation of the dark energy equation of state $w$ from $-1$ would have profound consequences for our understanding of dark energy, high precision on the relative velocity bias is required.  
 
In the present work, we use the 3PCF technique first proposed in Yoo, Dalal \& Seljak (2011) and developed further in SE15a to constrain the relative velocity bias to be $\leq 0.01$ for the SDSS BOSS DR12 CMASS galaxies (Eisenstein et al. 2011 for SDSS-III overview; Alam et al. 2015 for DR11 and DR12).  This precision is sufficient to ensure that the cosmic distance scale measurement from BOSS will not be systematically biased. It also suggests our technique is powerful enough to allow DESI to avoid this bias.  Our measurement is the most stringent constraint on the relative velocity bias available, and illustrates the power of the 3PCF for relative velocity constraints. 

The paper is laid out as follows.  In \S\ref{sec:bias_model}, we present our galaxy bias model including relative velocity bias. \S\ref{sec:results} describes our relative velocity constraint, discusses the other bias parameters, and reports the cosmic distance scale measured with this bias model.  We conclude in \S\ref{sec:concs}.

\section{Relative velocity bias model}
\label{sec:bias_model}
Following SE15a and Slepian \& Eisenstein (2016b; hereafter SE16b), the galaxy overdensity field $\deltagal$ traces the matter density field $\deltam$ and its square with two unknown bias coefficients, the linear bias $b_1$ and the non-linear bias $b_2$. The galaxy overdensity also traces the square of the local relative velocity with a third unknown bias coefficient, the relative velocity bias $b_v$. Note that at leading order in perturbation theory, the predicted 3PCF is the same whether one uses the Lagrangian or Eulerian relative velocity, as further discussed in S16b \S5. The bias model is
\begin{align}
\deltagal(\vx) &= b_1\deltam(\vx) + b_2\left[\deltam^2(\vx) - \left<\deltam^2(\vx) \right>\right]\\ \nonumber
& + b_v\left[v_{\rm s}^2(\vx)-1\right],
\label{eqn:bias_model}
\end{align}
where $v^2_{\rm s}(\vx)\equiv v_{\rm bc}^2(\vx)/\sigma_{\rm bc}^2$ is the relative velocity's square $v^2_{\rm bc}$ normalized by its mean square value $\sigma^2_{\rm bc}$.  $\deltam$ is the matter density field, which itself must be expanded to second-order in the linear density field $\delta$ as further discussed in SE15a and SE16b. This bias model does not include tidal tensor biasing, for which our companion paper (Slepian et al. 2016a; hereafter S16a) found mild evidence.  Future work may be incorporating a tidal tensor bias into the RV constraint; for now we note that the tidal tensor itself contributes broadband features to the 3PCF whereas the RV has a sharp, distinctive signature. We therefore do not expect adding a tidal tensor bias to substantially change the RV constraint.

Using the bias model (\ref{eqn:bias_model}), one can compute the 3PCF model including the relative velocity to lowest (fourth) order in the linear density field and including large-scale redshift-space distortions. Details of this computation and the resulting model are in SE16b.

We note that our 3PCF model requires as an input $\beta = f/b_1\approx \Omega_{\rm m}^{0.55}$, with $f=d\ln D/d\ln a$, $a$ the scale factor and $D$ the linear growth rate.  For reasons discussed in S16a \S6.2, we elect to fix $\beta$ at the beginning of our fitting,  meaning that after obtaining $b_1$ from the fitting we must check that $\beta$ is self-consistent.  Here we adopt two different $\beta$ values: for the data, $\beta = 0.44$, while for the mocks, $\beta = 0.40$, which is consistent with the fitted value of $b_1$ averaged over all mocks.

\section{Results}
\label{sec:results}
\subsection{Relative velocity constraint}
\label{subsec:constraint}
As described in S16a, we compute the 3PCF of the CMASS galaxies and the covariance matrix
in the Gaussian random field approximation.  We then fit to the 3PCF model based
on the bias model in \S\ref{sec:bias_model}. We marginalize over $\alpha$, the value of the BAO scale normalized to a fiducial sound horizon (it is unity if we input the correct cosmology), and over $c$, a free parameter describing deviations from the integral constraint and also intended to remove any survey-scale systematic bias.  The details of our fitting are further described in S16a.

Our fitted parameters are displayed in Table 1 (data) and Table 2 (mocks). We constrain the relative velocity bias as $b_v=-0.002\pm 0.01$.  These error bars are computed as the square root of the appropriate diagonal element of the bias covariance matrix as described in S16a.  

This result is highly consistent with results from two other methods of estimating the error bar. First, we know that the mock catalogs have $b_v\equiv 0$; no relative velocity biasing has been incorporated in them.  Our 3PCF fitting returns a mean value of $b_v=0$ where the mean is taken over the mocks, with a standard deviation of about $0.01$, as shown in Figure \ref{fig:velo_first_four}.  Second, we can compute $\sigma(b_v)$, the root mean square of the relative velocity bias marginalized over the integral constraint amplitude $c$ for each mock, as described for the other bias parameters in S16a.  We then find its average over all mocks as $\left<\sigma(b_v)\right>=0.97$.  Thus, all three error estimation methods concur that we can constrain the relative velocity bias with $0.01$ precision.

We note that the mean $b_v$ from the mocks is $\left<b_v\right>= -0.0031$, but that this is statistically distinguished from 0 given the $1/\sqrt{298}\approx 0.06$ error on the mean of the mocks.  This issue is similar to the possible systematic difference of the mean $\left<\alpha\right>$ over all mocks from unity in S16a. This value of $\left<b_v\right>$ for the mocks indicates that there is a small
discrepancy between the bias model and the mocks, which is causing
a systematic bias in this $b_v$ term. We will further  explore this point
with better 3PCF models and more extensive mock catalogs in future work.

\subsection{Other bias parameters}
\label{subsec:biases}
We now briefly discuss the other bias parameter values found for the mocks and for the data.  As Tables 1 and 2 show, the bias values and error bars we find are generally consistent between mocks and data. There is some disagreement between the error bar for $b_1$ from the data and from the scatter of the mocks.  We achieve a strong constraint on $b_1$, making a $1.10\%$ precision measurement.  We do not obtain a strong constraint on $b_2$; as discussed in SE16b and S16a, it is highly degenerate with $b_1$ and so will not be well-measured.  The integral constraint amplitude value  for the data is consistent with that we find for our minimal model (with $b_v\equiv 0$) in S16a, and this is also the case for the integral constraint amplitude for the mocks.  

Regarding the $\chi^2$, for the data we find $\chi^2/{\rm d.o.f} = 223.26/195$, indicating that our model fits the data fairly well. For the mocks, we find an average $\chi^2/{\rm d.o.f} = 195.34/195$, indicating that the mocks also are well-fit by the model. If the model truly describes the data or mocks, these $\chi^2$ have probabilities $P$ to occur by chance of respectively $P=0.08$ and $P=0.49$.

\begin{figure*}
\centering
\includegraphics[width=.485\textwidth]{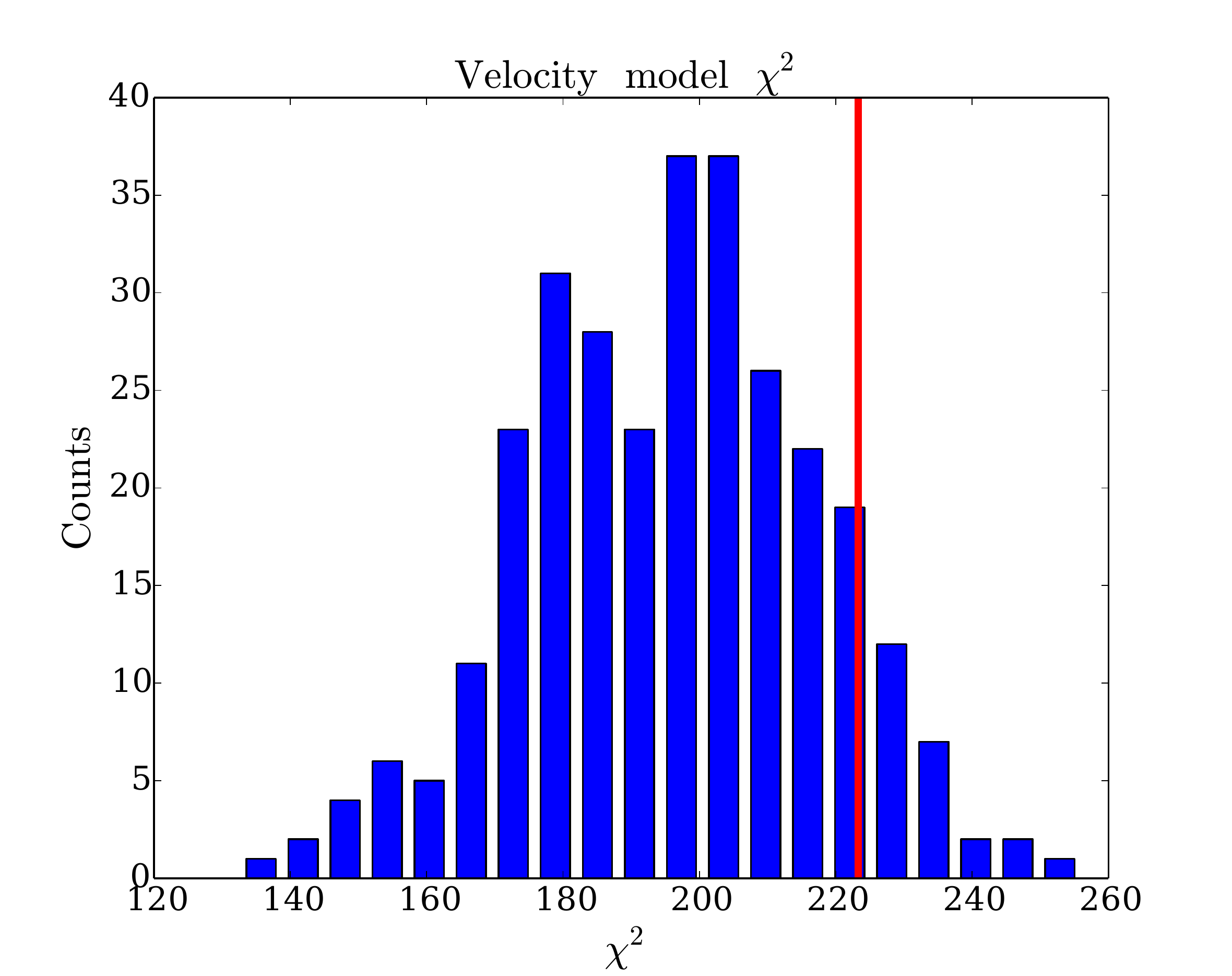}
\includegraphics[width=.495\textwidth]{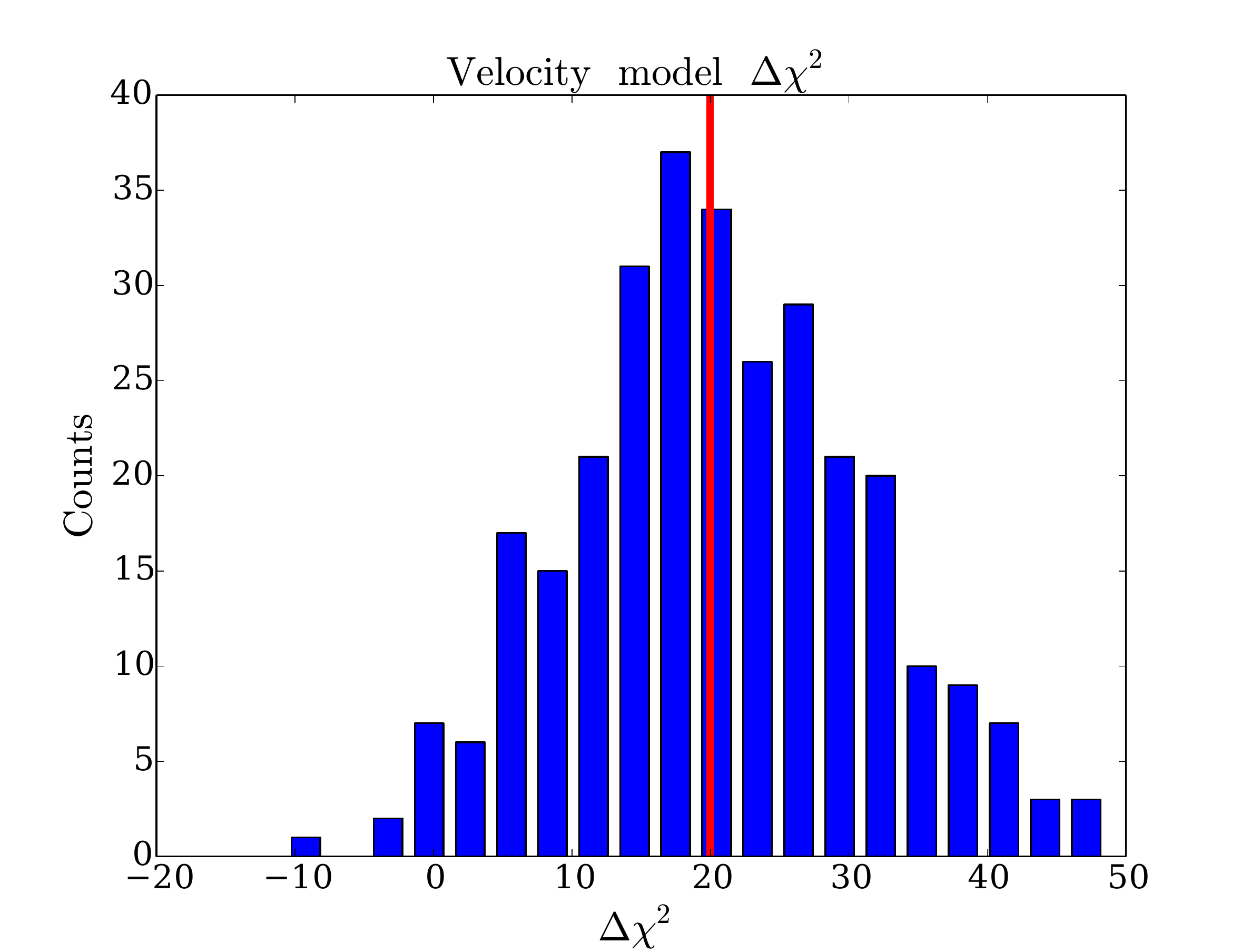}
\includegraphics[width=.495 \textwidth]{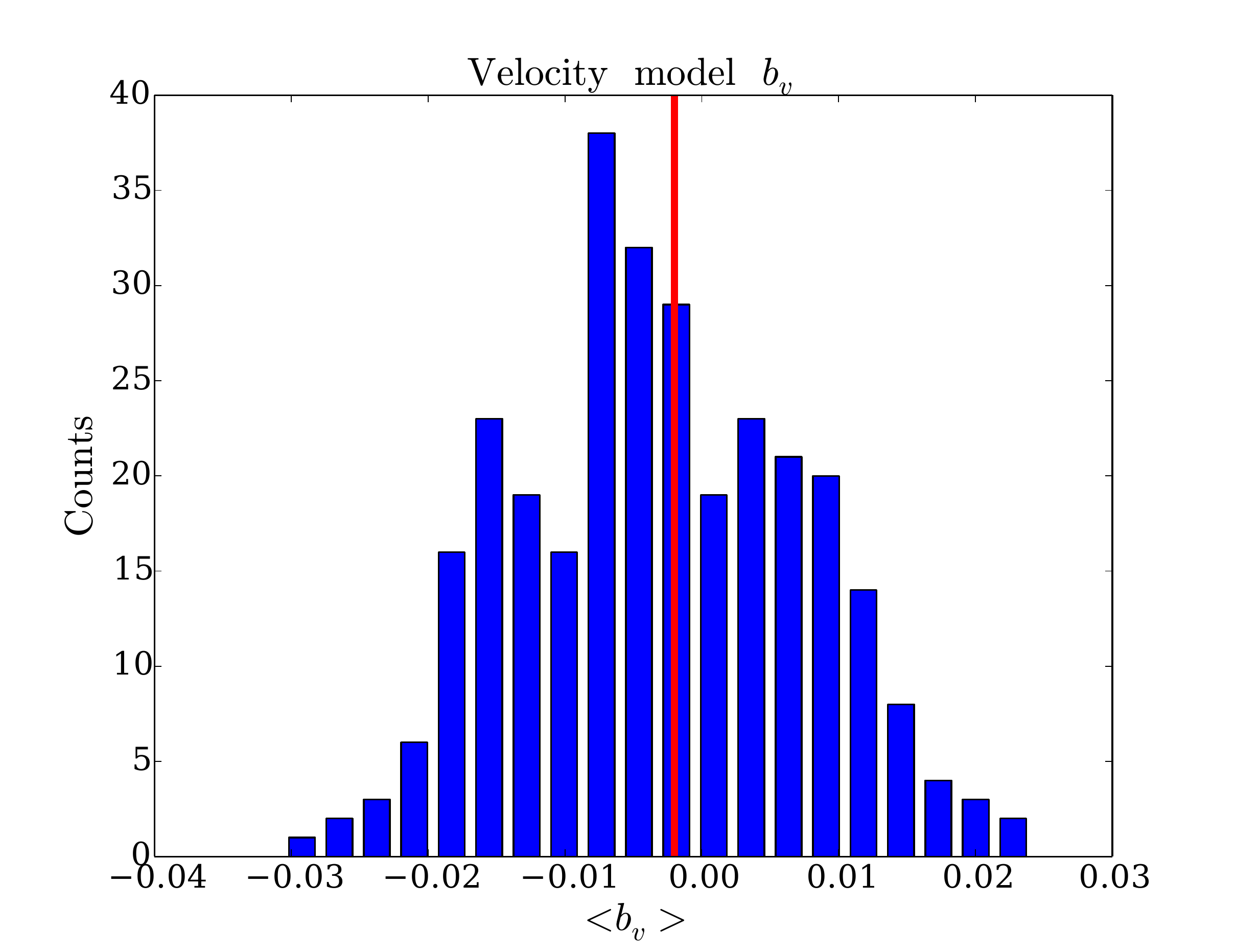}
\includegraphics[width=.495\textwidth]{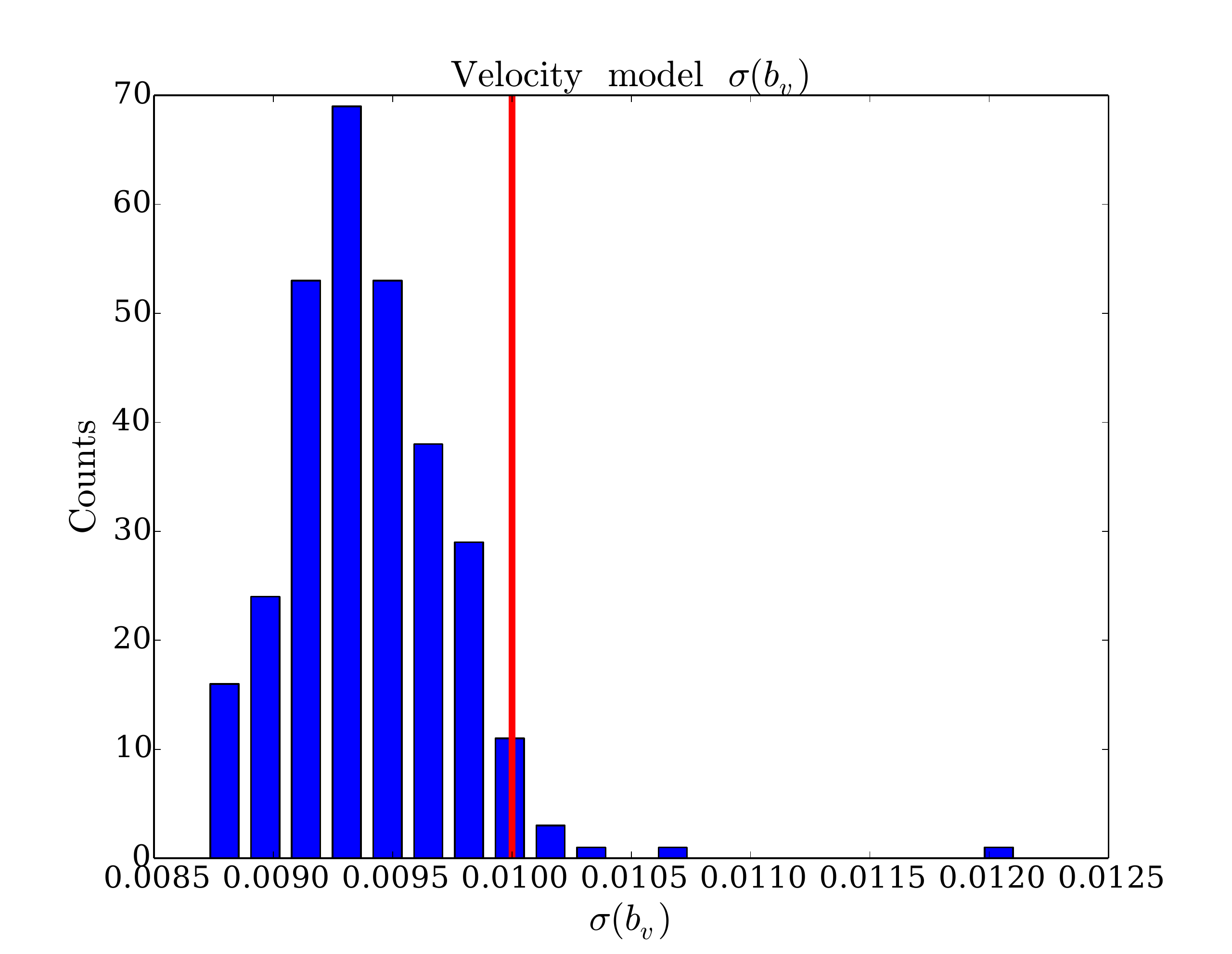}
\caption{The upper two panels show histograms of the best-fit $\chi^2$ and the $\Delta\chi^2$ with respect to the best-fit no-wiggle templates (here we set $b_v \equiv 0$ for the no-wiggle template) for the 298 \textsc{PATCHY} mocks (S16a and references therein). The red vertical line indicates the data values. These show that our goodness of fit and BAO significance are both fairly typical for a survey of this volume.  The bottom panels histogram the mock results for the relative velocity bias and the root mean square of this bias marginalized over the integral constraint amplitude. The true $b_v$ of all mocks is identically zero, so the $0.01$ scatter about zero in the left panel represents one estimate of our error bar on $b_v$. The estimate of the error as the root mean square of the mocks' $b_v$, shown in the lower right panel, again indicates $0.01$ precision on $b_v$.}
\label{fig:velo_first_four}
\end{figure*}

Figure \ref{fig:velo_first_four} shows the typicality of our results from the data with respect to the mocks. In each panel, the red line indicates the data value. The upper left panel shows that the $\chi^2$ for our best-fit model to the data is fairly typical for a survey of this size.  The upper right panel shows the difference in $\chi^2$ between the best-fit no-wiggle template (with $b_v \equiv 0$)  and the best-fit physical template (with BAO, and with $b_v$ free). Our BAO detection in this dataset is in the range we would expect for a survey of this size. 

The lower left panel of Figure \ref{fig:velo_first_four} shows the $b_v$ values measured for the mocks.  In truth, the mocks have $b_v = 0$, so any $b_v \neq 0$ we measure indicates the error bar on the $b_v$ constraint derived from the data. As expected, the center of the mocks' $b_v$ distribution is at $b_v = 0$, and the scatter is $0.01$.  As Table 1 indicates, for the data we find $b_v$ consistent with zero within our error bars. The right lower panel in Figure \ref{fig:velo_first_four} shows the root mean square $\sigma(b_v)$ computed from marginalizing $b_v$ and $b_v^2$ over the integral constraint amplitude, following the same procedure as outlined for the linear bias in S16a \S5.2.  The mean $\sigma(b_v)$ is about $0.0097$, and the data has $\sigma(b_v) = 0.010$; these values both indicate that we can constrain the relative velocity bias as $b_v<0.01$.

\begin{table}
\label{table:best_fit_params_data}
\begin{tabular}{|c|c|}
\hline
\;\;\;\;\; & \specialcell{Data: Velocity}\tabularnewline\hline
$\Delta\chi^2$&\specialcell{$19.99$}\tabularnewline\hline
$\alpha$& \specialcell{$0.990\pm0.020$}\tabularnewline\hline
$b_1$ &\specialcell{$1.776\pm0.020$}\tabularnewline\hline
$b_2$ &\specialcell{$0.52\pm0.17$}\tabularnewline\hline
$b_v$ & \specialcell{$-0.002\pm0.010$}\tabularnewline\hline
$c$ &\specialcell{$-0.014\pm0.003$}\tabularnewline\hline
$\chi^2$&\specialcell{$223.26$}\tabularnewline\hline
\end{tabular}
\centering
\caption{Table of best-fit parameters for the CMASS data.  $b_1$, $b_2$, and $b_v$ are the linear, non-linear, and relative velocity biases, and $c$ encodes the integral constraint.  $\Delta\chi^2$ describes the $\chi^2$ penalty a no-BAO model pays over a model with BAO. The value listed here implies  a $4.49\sigma$ BAO detection for the velocity model. $\alpha$ describes the inferred cosmic distance scale. The error bars quoted here are from the square root of the diagonal of the bias covariance matrix $\bf{C}_{\rm bias}$. Our error bar on the linear bias corresponds to $1.10\%$.}
\end{table}


\begin{table}
\label{table:best_fit_params_mocks}

\begin{tabular}{|c|c|}
\hline
\;\;\;\;\; & \specialcell{Mocks: Velocity}\tabularnewline\hline
$\Delta\chi^2$&\specialcell{$20.59$}\tabularnewline\hline
$\alpha$& \specialcell{$1.01\pm0.026\;(0.021)$}\tabularnewline\hline
$b_1$ &\specialcell{$1.936\pm0.030$}\tabularnewline\hline
$b_2$ &\specialcell{$0.50\pm0.21$}\tabularnewline\hline
$b_v$ & \specialcell{$-0.0031\pm0.0097$}\tabularnewline\hline
$c$ &\specialcell{$0.000\pm0.009$}\tabularnewline\hline
$\chi^2$&\specialcell{$195.34$}\tabularnewline\hline
\end{tabular}
\centering
\caption{Table of best-fit parameters for the mocks. We report the mean of each parameter over the 298 mocks. The error bars on these parameters are the standard deviation of the paramter taken over all mocks; for $\left<\alpha\right>$ we also report the average of the root mean square $\sigma(\alpha)$ over all mocks in parentheses. For the error bars on the biases we have held $\alpha$ fixed at its average value over all the mocks, as allowing $\alpha$ to float can artificially inflate the scatter in the biases. This point is further discussed in S16a \S9. Comparing the error bars here, from the scatter of the 298 mocks, to those reported in Table 1, mostly confirms that the error bars estimated from the bias covariance matrix are reasonable.  We further discuss this point in \S\ref{sec:results}.}
\end{table}

\begin{figure*}
\centering
\includegraphics[width=.49\textwidth]{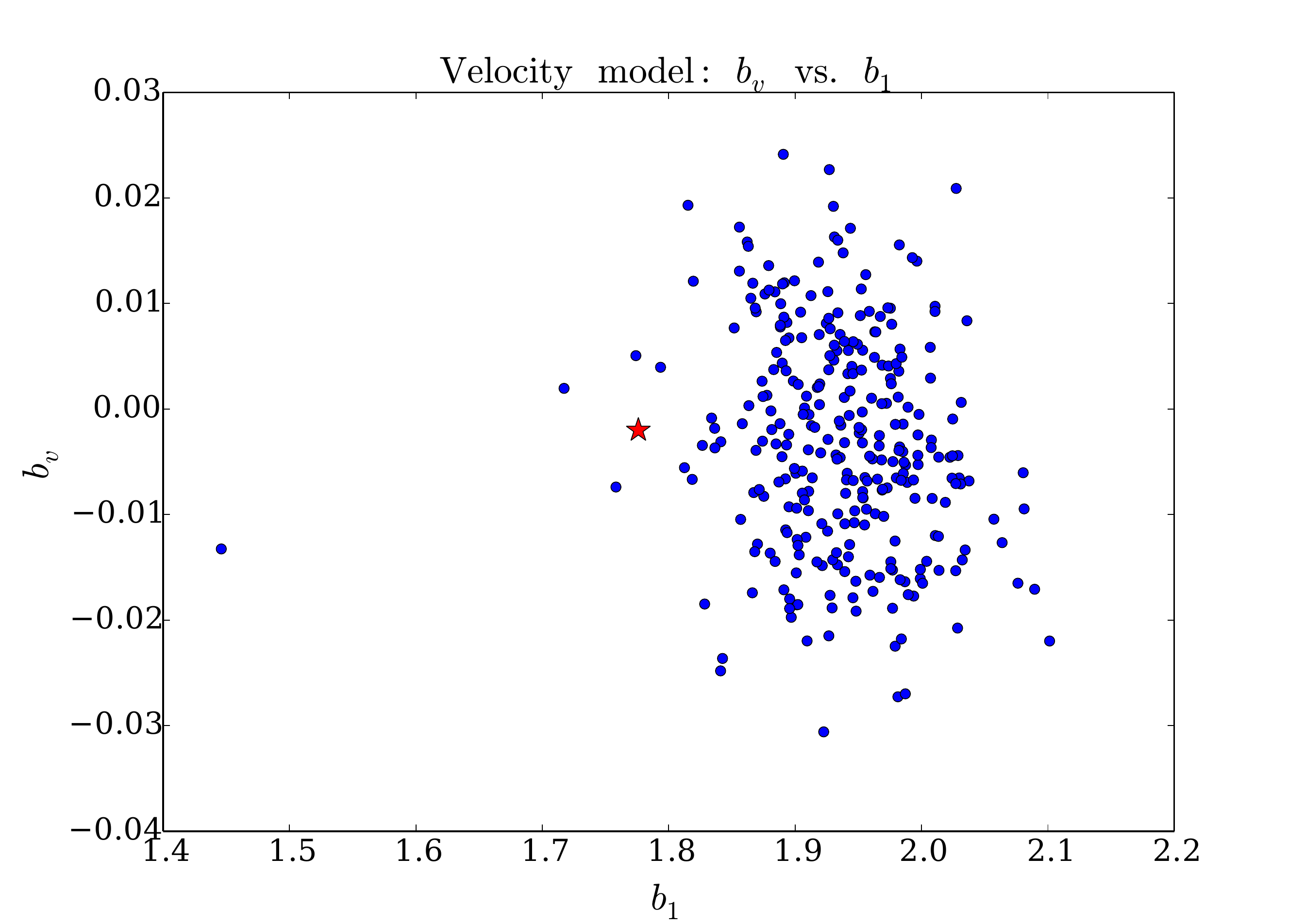}
\includegraphics[width=.486\textwidth]{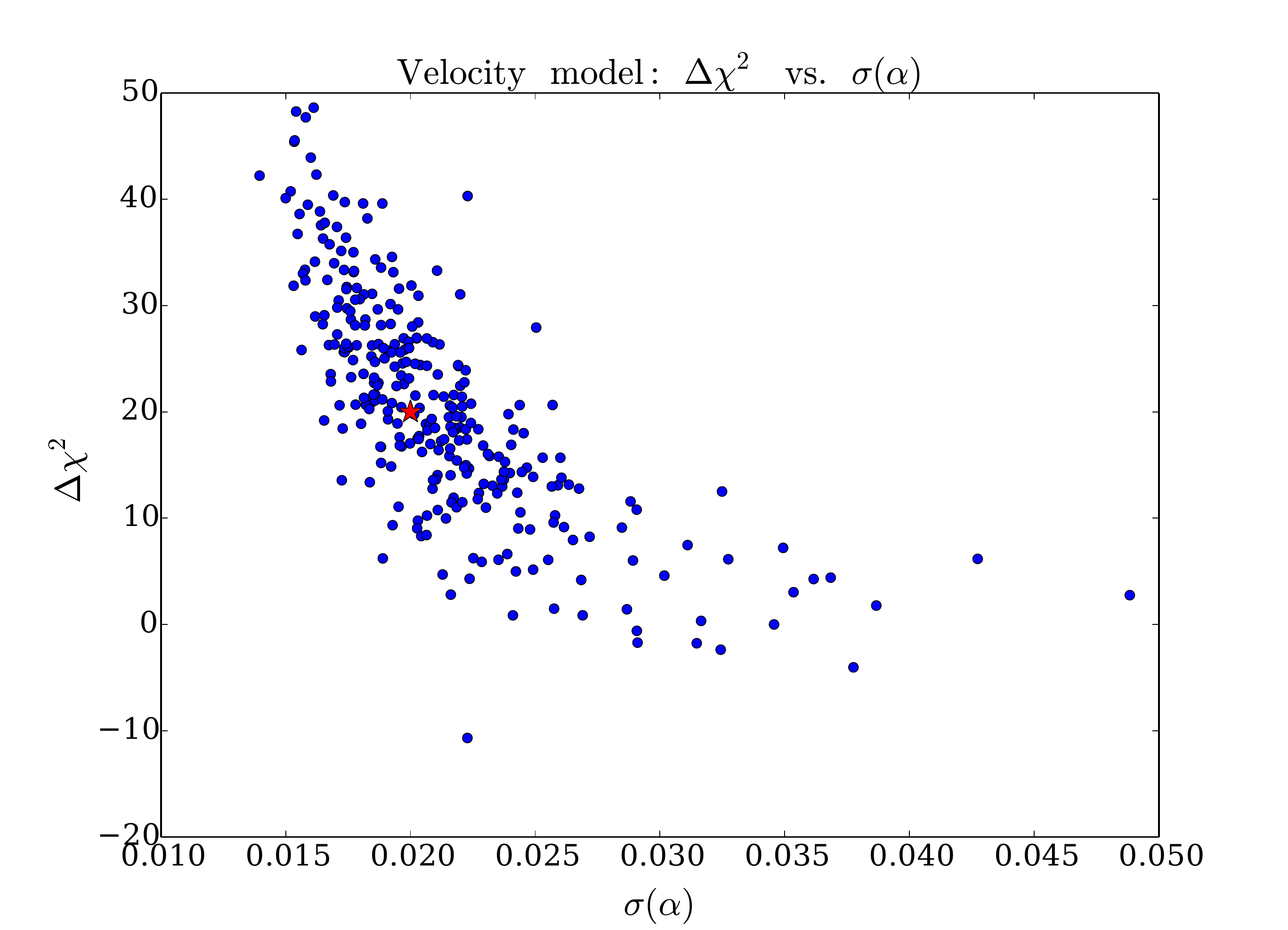}
\includegraphics[width=.49 \textwidth]{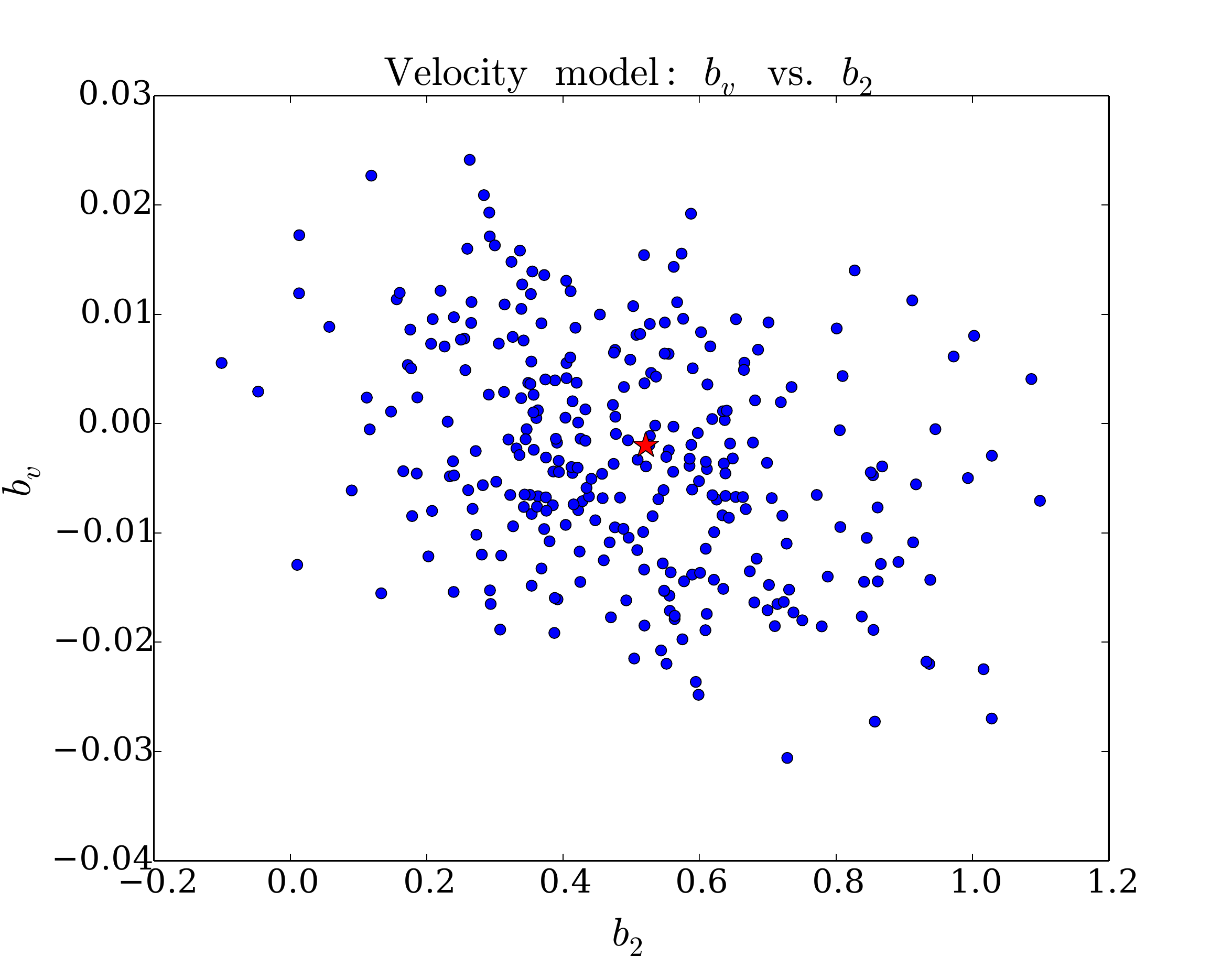}
\includegraphics[width=.49\textwidth]{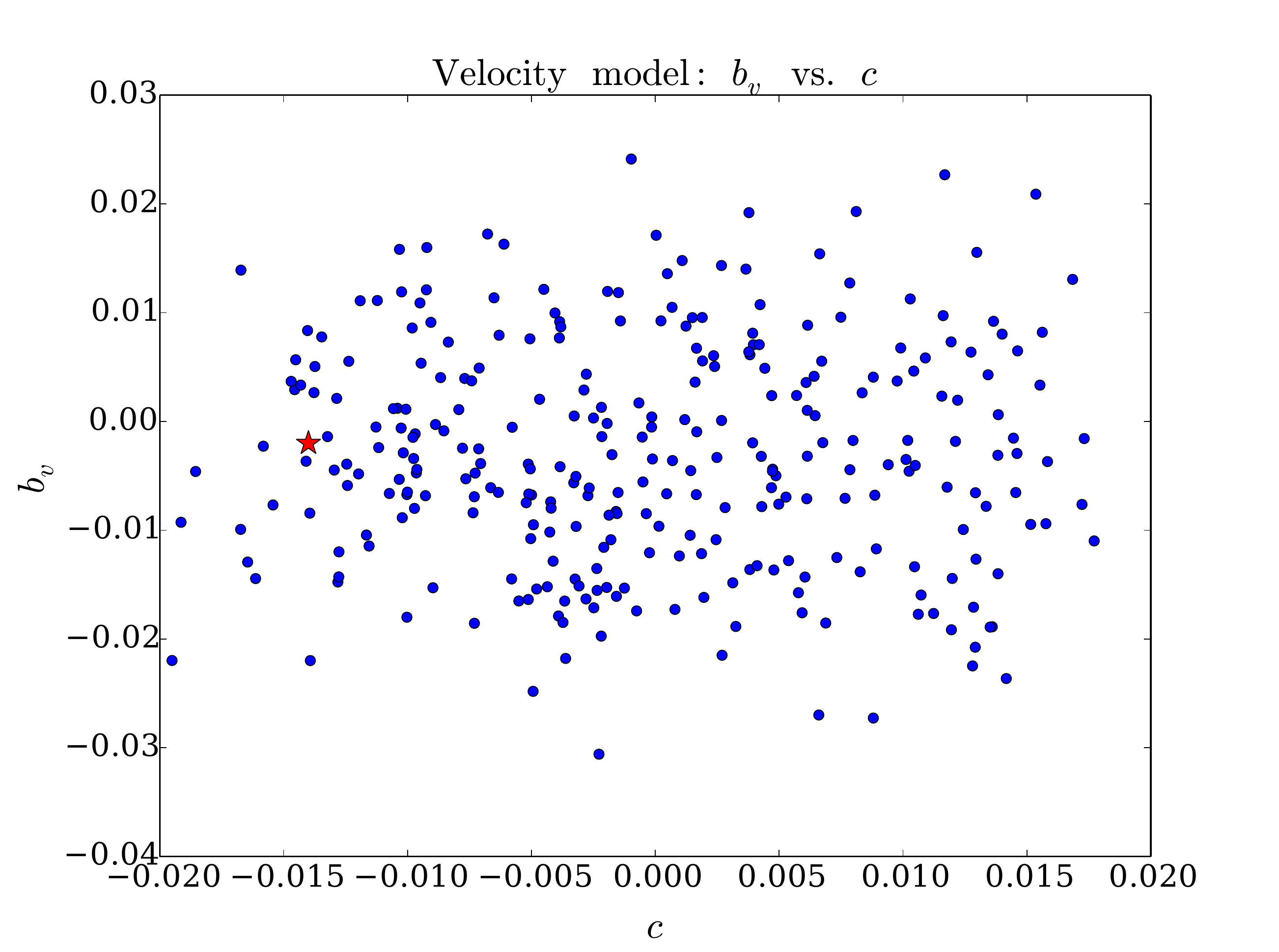}
\caption{Here we show several scatter plots illustrating the degeneracy structure of $b_v$ with respect to the other parameters of our fits. In all panels the data values are marked by a red star. The upper left panel shows that $b_v$ is not highly degenerate with $b_1$. The lower left panel shows there is an extremely mild anti-correlation between $b_2$ and $b_v$. The lower right panel shows that there is no correlation between the integral constraint amplitude $c$ and $b_v$. The upper right panel shows that as the significance of our BAO detection rises, the root mean square error on $\alpha$ improves, as expected.}
\label{fig:scatter_biases}
\end{figure*}

\subsection{Cosmic distance scale}

To convert $\alpha$ into a physical distance scale $\DV$ to redshift $0.57$, we generalize the formula for $\DV$ of Anderson et al. (2014) to varying $\omegam$ and redshift; we also convert to the \textsc{PATCHY} cosmology (S16a and references therein) and from $\Mpch$ to ${\rm Mpc}$. We find
\begin{align}
\DV = \alpha \times 2054.4\;{\rm Mpc}\left(\frac{r_{\rm d}}{r_{\rm d,\;\textsc{PATCHY}}}\right)
\end{align}
where $r_{\rm d}$ is the sound horizon at decoupling and $r_{\rm d,\;\textsc{PATCHY}}$ is the sound horizon at decoupling for the PATCHY cosmology. We thus find
\begin{align}
D_{\rm V,\; velocity}(z_{\rm survey})  =\;& 2034\pm 41\;{\rm Mpc\;(stat)}\pm 20\;{\rm Mpc\;(sys)}\nonumber\\
&\times \left(\frac{r_{\rm d}}{r_{\rm d,\;\textsc{PATCHY}}}\right).
\end{align}
From fitting the 2PCF of SDSS DR11 including reconstruction, Anderson et al. (2014) found 
\begin{align}
&D_{\rm V,\; Anderson} (z_{\rm survey}) = 2034\pm 20\;{\rm Mpc}\left(\frac{r_{\rm d}}{r_{\rm d,\;{\textsc PATCHY}}}\right)
\end{align}
while from the SDSS DR12 CMASS 2PCF including reconstruction, Cuesta et al. (2016) found
\begin{align}
&D_{\rm V,\; Cuesta}(z_{\rm survey})  = 2036\pm 21\;{\rm Mpc}\left(\frac{r_{\rm d}}{r_{\rm d,\;{\textsc PATCHY}}}\right).
\end{align}
From the reconstructed multipoles of the CMASS DR12 power spectrum, Gil-Mar\'in et al. (2016) found 
\begin{align}
&D_{\rm V,\; Gil-Marin}(z_{\rm survey})  = 2023\pm 18\;{\rm Mpc}\left(\frac{r_{\rm d}}{r_{\rm d,\;{\textsc PATCHY}}}\right)
\end{align}
We have adjusted the measured $\DV$'s of these works appropriately to be quoted in terms of our fiducial \textsc{Patchy} sound horizon.

Our velocity model measurements are therefore highly consistent with the latest 2PCF and power spectrum BAO analysis results. None of the other works we quote incorporated a relative velocity bias in their fitting, but given that our data prefers a $b_v$ that is nearly zero, we indeed expect our velocity model distance scale measurement to be consistent with theirs. Our slightly larger error bars reflect that we achieve a precision of roughly $2.0\%$ on the distance scale while the 2PCF or power spectrum measurements achieve a precision of roughly $1.0\%$. Our results are also consistent with the measured distance scale in the final cosmological analysis of the SDSS DR12 combined sample (Alam et al. 2016).

\section{Conclusions}
\label{sec:concs}
We have shown that the 3PCF permits a $0.01$ precision measurement of the relative velocity bias, translating to about $0.5\%$ of the linear bias.  We have shown three different estimates of this error bar that are all consistent with each other.  For the data we find $b_v$ consistent with zero within our error bars.  

The constraint of Yoo \& Seljak (2013) from the power spectrum used $260,000$ SDSS DR11 galaxies to place the constraint $b_{v,{\rm YS}} < 0.033$.  In their bias model, the relative velocity's square is normalized by its 1-D variance, $\sigma_{{\rm bc},\; 1-{\rm D}}^2$, which is $1/3$ the 3-D variance $\sigma_{\rm bc}^2$ used in our bias model (\ref{eqn:bias_model}). Holding the combination $b_v(\vbc^2/\sigma_{\rm bc}^2)$ constant, as this is what enters the bias models, the normalization difference means that their measured $b_v$ should be multiplied by a factor of $3$ to be compared with ours. Thus as we define $b_v$ the Yoo \& Seljak (2013) constraint is $b_v < 0.1$. The constraint $b_v<0.01$ of this work is a factor of ten tighter.

Were our 3PCF technique equally good as the power spectrum analysis, we would expect a $0.58$ precision constraint (the precisions simply scale as $\sqrt{N_{\rm g}}$, with $N_{\rm g}$ the number of galaxies).  Finding a $0.01$ constraint thus shows the superiority of the 3PCF for these measurements by roughly a factor of six.  

The relative velocity effect can bias the BAO scale measured in the 2PCF, and thus a tight constraint on $b_v$ is essential for present surveys such as BOSS and future efforts like DESI to remain unbiased.  The constraint we find in this work is tight enough that the shift in $\alpha$ measured from the 2PCF will be less than $0.3\%$, using a recalculation of Blazek, McEwen \& Hirata (2016) Figure 2 for the appropriate linear bias and survey redshift for CMASSS to translate $b_v/b_1$ to a shift in $\alpha$. The BOSS survey can thus control $b_v$ to a level equivalent to $\sim 1/3$ of its BAO
statistical precision.  For surveys of larger volume at similar number density,
this indicates that the RV effect can be sufficiently controlled.

In closing, we highlight that the consistency with zero of our measured $b_v$ for the CMASS data has interesting possible implications for galaxy formation models. Our constraint on $b_v$ suggests that galaxies do not have strong memories of their less-massive high redshift progenitors. Given that only a small fraction of the stars in LRGs at $z\sim 0$ were produced in the high-redshift small halos most affected by the relative velocity, this finding is not unexpected. Our constraint suggests that feedback is likely efficient at erasing any differences between galaxies formed in high relative velocity regions and low relative velocity regions.  While mergers also play a role in the evolution of small high-redshift halos into the LRGs used for the BAO, the relative velocity's coherence scale is sufficiently large that we do not expect mergers could by themselves erase a relative velocity imprint. Further exploration of this point may be a worthwhile avenue of future work.


\section*{Acknowledgements}
We thank Blakesley Burkhart, Cora Dvorkin, Douglas Finkbeiner, Margaret Geller, Abraham Loeb, Philip Mocz, Ramesh Narayan, Stephen Portillo, Roman Scoccimarro, Uro\v{s} Seljak, Joshua Suresh, Licia Verde, and Martin White for useful conservations. This material is based upon work supported by the National Science Foundation Graduate Research Fellowship under Grant No. DGE-1144152; DJE is supported by grant DE-SC0013718 from the U.S. Department of Energy.

This material is based upon work supported by the National Science Foundation Graduate Research Fellowship under Grant No. DGE-1144152; DJE is supported by grant DE-SC0013718 from the U.S. Department of Energy. JB is supported by a CCAPP Fellowship. HGM acknowledges Labex ILP (reference ANR-10-LABX-63) part of the Idex SUPER, and received financial state aid managed by the Agence Nationale de la Recherche, as part of the programme Investissements d'avenir under the reference ANR-11-IDEX-0004-02. SH is supported by NSF AST1412966, NASA -EUCLID11-0004 and NSF AST1517593 for this work. WJP acknowledges support from the UK Science and Technology Facilities
Research Council through grants ST/M001709/1 and ST/N000668/1, the
European Research Council through grant 614030 Darksurvey, and the UK
Space Agency through grant ST/N00180X/1. GR acknowledges support from the National Research Foundation of Korea (NRF) through NRF-SGER 2014055950 funded by the Korean Ministry of Education, Science and Technology (MoEST), and from the faculty research fund of Sejong University in 2016. FSK thanks support from the Leibniz Society for the Karl-Schwarzschild fellowship. 

Funding for SDSS-III has been provided by the Alfred P. Sloan Foundation, the Participating Institutions, the National Science Foundation, and the U.S. Department of Energy Office of Science. The SDSS-III web site is http://www.sdss3.org/.

SDSS-III is managed by the Astrophysical Research Consortium for the Participating Institutions of the SDSS-III Collaboration including the University of Arizona, the Brazilian Participation Group, Brookhaven National Laboratory, Carnegie Mellon University, University of Florida, the French Participation Group, the German Participation Group, Harvard University, the Instituto de Astrofisica de Canarias, the Michigan State/Notre Dame/JINA Participation Group, Johns Hopkins University, Lawrence Berkeley National Laboratory, Max Planck Institute for Astrophysics, Max Planck Institute for Extraterrestrial Physics, New Mexico State University, New York University, Ohio State University, Pennsylvania State University, University of Portsmouth, Princeton University, the Spanish Participation Group, University of Tokyo, University of Utah, Vanderbilt University, University of Virginia, University of Washington, and Yale University.

\section*{References}
\hangindent=1.5em 
\hangafter=1
\noindent Alam S. et al., 2015, ApJS, 219, 12

\hangindent=1.5em 
\hangafter=1
\noindent Alam S. et al., 2016, preprint (arXiv:1607.03155).

\hangindent=1.5em 
\hangafter=1
\noindent Beutler F., Blake C., Koda J., Mar\'in F., Seo H.-J., Cuesta A.J. \& Schneider D., 2016, MNRAS 455, 3, 3230-3248.

\hangindent=1.5em
\hangafter=1
\noindent Blazek J., McEwen J. \& Hirata C., 2016, PRL 116, 12, 121303.

\noindent Bond J.R. \& Efstathiou G., 1984, ApJ 285, L45.

\noindent Bond J.R. \& Efstathiou G., 1987, MNRAS 226, 655-687.

\hangindent=1.5em
\hangafter=1
\noindent Cuesta A.J. et al., 2016, MNRAS 457, 2, 1770-1785. 

\hangindent=1.5em 
\hangafter=1
\noindent Eisenstein D.J. et al. 2011, AJ, 142, 72

\hangindent=1.5em
\hangafter=1
\noindent Eisenstein D.J. \& Hu W., 1998, ApJ 496, 605. 

\hangindent=1.5em
\hangafter=1
\noindent Eisenstein D.J., Seo H.-J. \& White M., 2007, ApJ 664, 2, 660-674.

\hangindent=1.5em
\hangafter=1
\noindent Gil-Mar\'in H. et al., 2016, MNRAS, doi:10.1093/mnras/stw1264.

\hangindent=1.5em
\hangafter=1
\noindent Holtzmann J.A., 1989, ApJS 71, 1.

\noindent Hu W. \& Sugiyama N., 1996, ApJ 471:542-570.

\noindent Peebles P.J.E. \& Yu J.T., 1970, ApJ 162, 815.

\hangindent=1.5em
\hangafter=1
\noindent Sakharov A.D., 1966, Soviet Journal of Experimental and Theoretical Physics 22, 241.

\hangindent=1em
\hangafter=1
\noindent Slepian Z. \& Eisenstein D.J., 2015a, MNRAS 448, 1, 9-26.

\hangindent=1em
\hangafter=1
\noindent Slepian Z. \& Eisenstein D.J., 2015b, MNRAS 454, 4, 4142-4158.

\hangindent=1em
\hangafter=1
\noindent Slepian Z. \& Eisenstein D.J., 2015c, MNRASL 455, 1, L31-L35.

\hangindent=1em
\hangafter=1
\noindent Slepian Z. \& Eisenstein D.J., 2016a, MNRAS 457, 24-37. 

\hangindent=1em
\hangafter=1
\noindent Slepian Z. \& Eisenstein D.J., 2016b, preprint (arXiv:1607.03109). 

\hangindent=1em
\hangafter=1
\noindent Slepian Z.  et al., 2015, preprint (arXiv:1512.02231). 

\hangindent=1em
\hangafter=1
\noindent Slepian Z.  et al., 2016a, preprint (arXiv:--). 

\noindent Sunyaev R.A. \& Zel'dovich Ya. B., 1970, Ap\&SS 7, 3.

\hangindent=1.5em
\hangafter=1
\noindent Tseliakhovich,D. \& Hirata C. 2010, PRD, 82, 083520.

\hangindent=1.5em
\hangafter=1
\noindent Yoo J., Dalal N. \& Seljak U., 2011, JCAP, 7, 018.

\hangindent=1.5em
\hangafter=1
\noindent Yoo J. \& Seljak U., 2013, PRD 88, 10, 103520.

\end{document}